\documentclass[12pt]{article}
 
\def\ltsim{\lower3pt\hbox{$\, \buildrel < \over \sim \, $}}
\def\gtsim{\lower3pt\hbox{$\, \buildrel > \over \sim \, $}}
\def\be{\begin{equation}}
\def\ee{\end{equation}}
\def\ba{\begin{eqnarray}}
\def\ea{\end{eqnarray}}
\def\ga{\mathrel{\raise.3ex\hbox{$>$\kern-.75em\lower1ex\hbox{$\sim$}}}}
\def\la{\mathrel{\raise.3ex\hbox{$<$\kern-.75em\lower1ex\hbox{$\sim$}}}}

\def\bo{ { \sqcup\llap{ $\sqcap$} } }
\begin{document}

\baselineskip=16pt 
\begin{titlepage}
\rightline{OUTP-00-47P}
\rightline{hep-th/0011138}
\rightline{November 2000}  
\begin{center}

\vspace{0.5cm}

\large {\bf The $m\rightarrow0$ limit for massive graviton in $dS_4$
and $AdS_{4}$  \\
 How to circumvent the van Dam-Veltman-Zakharov discontinuity}

\vspace*{5mm}
\normalsize

{\bf Ian I. Kogan\footnote{i.kogan@physics.ox.ac.uk}, Stavros
Mouslopoulos\footnote{s.mouslopoulos@physics.ox.ac.uk} and Antonios Papazoglou\footnote{a.papazoglou@physics.ox.ac.uk}}

\smallskip 
\medskip 
{\it Theoretical Physics, Department of Physics, Oxford University}\\
{\it 1 Keble Road, Oxford, OX1 3NP,  UK}
\smallskip

\vskip0.6in \end{center}
 
\centerline{\large\bf Abstract}

We show that, by considering physics in $dS_{4}$ or $AdS_{4}$ spacetime, 
one can circumvent the van Dam - Veltman - Zakharov 
theorem which requires that the extra polarization states  of a massive
graviton do not decouple in the massless limit. It is shown that the smoothness of the $m\rightarrow{0}$
limit is ensured if the H (``Hubble'') parameter, associated with the
horizon of the $dS_4$ or $AdS_{4}$ space, tends to zero slower than the mass of the graviton $m$.

\vspace*{2mm} 

\end{titlepage}

\section{Introduction}

Recently, there has been a lot of activity on models that suggest that
a part \cite{bigravity,KR,multi} or all \cite{GRS,KR,multi}
of gravitational interactions come from massive gravitons. The massive
gravitons in these models occur as the result of the dimensional reduction
of a theory of gravity in more that four dimensions, something
 well motivated from String Theory. 

In the first kind of model \cite{bigravity,KR,multi} apart of the massless graviton, there
exist ultralight Kaluza-Klein (KK) state(s) that lead to  a ``multigravity'' scenario, in the sense that
gravitational interactions  are due to the net effect of the
massless graviton and the ultralight state(s). In this case, the Cavendish
bounds on gravitational interactions are satisfied, since it can be
arranged that the rest of the KK tower is much heavier and contributes well below the
submillimeter region. In this scenario, modifications to gravity at large
scales will appear as we probe distances of the order of the Compton
wavelength of the ultralight KK state(s). The phenomenological signature of
this will be that gravitational interactions will be reduced or almost
switched off (depending on the choice of parameters of the model) at
ultralarge scales.

In the second kind of model \cite{GRS,KR,multi}, there is no
normalizable massless mode and 4D gravity at intermediate scales is
reproduced from a resonance-like behavior \cite{reso,dvali1} of the wavefunctions of the
KK states continuum. In other words 4D gravity in this case is
effectively reproduced from a small band of KK states starting from zero mass. In this
picture modifications of gravity will begin at scales that 
correspond to the width of the resonance that appears in the form of
the coupling of these states to matter. The phenomenological signature
of these modifications will be that the four dimensional Newton's Law
(\textit{i.e.} inverse square)
will  change to a five dimensional one (\textit{i.e.} inverse cube)
at ultralarge distances. In both kind of models, these modifications
 can be confronted with current observations of the CBM
power spectrum \cite{cmb} and are consistent with the data at present.

However, one should be careful when dealing with models with
ultralight massive KK states because it is known  that 
 the extra polarizations of the massive gravitons do not
decouple in the limit of vanishing mass, the famous van Dam
- Veltman - Zakharov \cite{VZ}
discontinuity. This could make these models disagree \cite{dvali1}
with standard tests of General Relativity, as for example the bending of
the light by the sun. Furthermore, the moduli (radions) associated with the perturbations of the
$''-''$ branes are necessarily physical ghost fields \cite{ghost}, therefore
unacceptable. The latter problem is connected to the violation of the
weaker energy condition \cite{weak} on $''-''$ branes sandwiched between $''+''$
 branes.  In the GRS model this radion cancels \cite{rest} the extra
polarizations of the massive gravitons and gives the graviton
propagator the correct tensorial structure at intermediate
distances. However, the model has still an explicit ghost in the
spectrum which reveals itself as scalar antigravity at cosmological
scales. A mechanism of
cancelling  \textit{both} the extra massive graviton polarizations \textit{and} the
radion field contribution was suggested in \cite{KR,multi} and
involves some bulk dynamics which are necessary to stabilize
the system, based on a scenario described in \cite{olive}. This mechanism is however non-local in the extra
dimension and because of this may not be very attractive.


In the present paper we will demonstrate that there is actually a way
out of the first problem. The second
problem can be avoided by considering models with only positive
tension branes but this will be addressed in another publication
\cite{++}.  Here we demonstrate that due to an unusual
property of the graviton propagators in $dS_{4}$ or $AdS_{4}$ spacetime, we are able
to circumvent the  van Dam - Veltman - Zakharov no go theorem
(actually the $dS_{4}$ was considered in \cite{HIG} using a different method). In more
detail, it is known that in flat spacetime the $m\rightarrow0$ limit
of the massive graviton propagator does not give the massless one due
to the non-decoupling of the additional longitudinal components. This
generates  the well known discontinuity between massive and massless states. Considering the massive graviton
propagator in $dS_{4}$ or $AdS_{4}$ spacetime we can show that this result
persists if $m/H \rightarrow \infty$ where $H$ is the  ``Hubble'' parameter,  \textit{i.e.} the
discontinuity is still present in the  $m\rightarrow 0$ limit if it
happens that 
$m/H \rightarrow \infty$. However,
in the case that  $m/H \rightarrow 0$, we will explicitly show that
the $m \rightarrow 0$ limit is smooth. This
 is an important result since it gives us the possibility to circumvent
the van Dam - Veltman - Zakharov no go theorem about the
non-decoupling of the  extra graviton
polarizations. Thus, in the limit that $m/H \rightarrow 0$ 
all the extra polarizations of the graviton decouple, giving an
effective theory with  massless graviton with just two polarization
states.

Here we
have to make an important comment about the relationship of the parameters
$m$, $H$. From a four dimensional point of view these parameters are
independent. However, the models that give the additional KK contributions of
massive gravitons to gravity are higher dimensional models. After the
dimensional reduction, they give us an effective four
dimensional Lagrangian in which in general $m$ and $H$
depend on common parameters (\textit{e.g.} the effective four dimensional
cosmological constant).  The behaviour of $m/H$  is  model dependent
and explicit models  that
satisfy the smoothness requirement as $m/H \rightarrow 0$  can be
found. An interesting example will be given elsewhere \cite{++}.

Furthermore, in such models if we keep $m/H$ finite but small, the
extra graviton polarization states couple more weakly to matter
than the transverse states. This gives us the possibility that we can
have a model with massive gravitons  that do not violate the
observational bounds of \textit{e.g.} the bending of light by the
sun. Moreover, such models predict modifications of gravity at all
scales which could be measured by higher precision observations. However, in such a case 
even though the above can make ``multigravity'' models viable and interesting,  the condition that the
mass of the massive graviton must always scale faster that the ``Hubble'' parameter
 implies that the dramatic long distance effects of modifications of
gravity (\textit{e.g.} reduction of Newton's constant or transition
to 5D law)
will not reveal themselves until super-horizon scales. Thus, in this case the
horizon acts as a curtain that prevents the long distance modifications
of gravity due to the massive KK mode(s) to be observable.

\section{Graviton propagator in flat spacetime}

In order to understand how we can circumvent the van
Dam-Veltman-Zakharov discontinuity, it is useful to review the
forms of the massive and massless graviton propagators in flat spacetime
and the resultant phenomenological differences.

The celebrated  van
Dam - Veltman - Zakharov discontinuity is evident from the different form
of the propagators that correspond to the massive and massless graviton.
In  more detail, the form of the massless graviton propagator in flat
spacetime (in momentum space) has the form: 
\be
G^{\mu\nu;\alpha\beta}=\frac{1}{2}~ \frac{\left(\eta^{\mu\alpha}\eta^{\nu\beta} +
\eta^{\nu\alpha}\eta^{\mu\beta}\right) - \eta^{\mu\nu}\eta^{\alpha\beta} 
  }{p^2 }+ \cdots
\ee
where we have omitted terms that do not contribute when contracted
with a conserved $T_{\mu \nu}$.

On the other hand, the 
four-dimensional  massive graviton propagator (in momentum space) has the form: 

\be
G^{\mu\nu;\alpha\beta}=\frac{1}{2} ~ \frac{\left(\eta^{\mu\alpha}\eta^{\nu\beta} +
\eta^{\nu\alpha}\eta^{\mu\beta}\right) - \frac{2}{3}~ \eta^{\mu\nu}\eta^{\alpha\beta} 
 }{p^2 -m^2}+\cdots
\ee

 This discontinuity has observable phenomenological implications in
standard tests of Einsteinian gravity and particularly in
the  bending of light by the Sun. For example, if
gravity is due to the exchange of a massive spin 2 particle,
then the deflection angle of light would be $25\%$ smaller than if it
corresponds to the  exchange of the massless graviton.  
The fact that the  bending of the light by the sun agrees with 
the prediction of  Einstein's theory  to $1\%$  accuracy,  rules out
the possibility that gravity is due to massive graviton exchange
irrespective  of
how small the mass is\footnote{Of course there remains the possibility 
that a small
fraction of the gravitational interactions are associated with a
massive graviton component in the presence of a dominant massless
graviton component. 
 This can be realized by having an ultralight spin-2 particle with a
very small coupling compared to graviton's one \cite{bigravity}.}.

\section{Graviton propagator in $dS_4$ and $AdS_4$ space}

In this section we will present the forms of the massless and massive
graviton propagators in the case of $dS_4$ and $AdS_4$ spacetime with arbitrary
``Hubble'' parameter $H$ and graviton mass $m$. Our purpose is to examine the behaviour of
these propagators in the limit where $H\rightarrow0$ and the limit
where the mass of the graviton tends to zero.
 For simplicity we will do our calculations in Euclidean $dS_4$ or
$AdS_4$ space. We can use for metric the one of the stereographic
projection of the sphere or the hyperboloid\footnote{Note that \cite{mless} and
\cite{mive} whose results we use in the following have a different
metric convention, but this makes no difference for our calculations.}:
\vspace{0.5cm}
\be
ds^2= \frac{\delta_{\mu\nu}}{\left(1 \mp {H^2 x^2 \over4}\right)^2}dx^{\mu}dx^{\nu}\equiv g^{0}_{\mu\nu}dx^{\mu}dx^{\nu}
\ee
where $x^2=\delta_{\mu\nu}x^{\mu}x^{\nu}$ and the scalar curvature is $R=\mp 12~H^2$. From now on, the upper
sign corresponds to $AdS_4$ space while the lower to $dS_4$ space.
The fundamental invariant in these spaces is the geodesic distance $\mu(x,y)$
between two points $x$ and $y$. For convenience, we will introduce
another invariant $u$ which is related with the geodesic distance 
by the relation $u=\cosh (H \mu)-1$ for $AdS_4$ ~($u\in[0,\infty)$)
and the relation $u=\cos (H \mu)-1$ for $dS_4$ ~($u\in[-2,0]$).  In
the small distance limit $u\sim \pm {\mu^2 H^2 \over 2}$.

This background metric is taken by the the Einstein-Hilbert action:
\be
S=\int d^4x \sqrt{g}\left(2M^2 R-\Lambda\right)
\ee
where the cosmological constant is
$\Lambda=\mp 12~H^2 M^2$ and M the 4D fundamental scale. The spin-2 massless 
graviton field can be obtained by the linear metric
fluctuations $ds^2=\left(g^0_{\mu\nu}+h_{\mu\nu}\right)dx^{\mu}dx^{\nu}$. This
procedure gives us the analog of the Pauli-Fierz graviton action in
curved space:
\ba
\frac{S_0}{2 M^2}= \int d^4x \sqrt{g^0} \left\{-{1\over 4}h \bo^0 h +{1\over
2}h^{\mu \nu}\nabla^0_{\mu}\nabla^0_{\nu}h +{1\over 4}h^{\mu \nu} \bo^0
h_{\mu \nu} - {1\over
2} h^{\mu \nu}\nabla^0_{\nu}\nabla^0_{\kappa}
h^{\kappa}_{\mu}\right.\cr \pm \left. {1\over
2}H^2\left({1\over 2}h^2 +h_{\mu \nu}h^{\mu \nu}\right)\right\} 
\ea
The above action is invariant under the gauge transformation $\delta
h_{\mu\nu}=\nabla^0_{\mu}\xi_{\nu}+\nabla^0_{\nu}\xi_{\mu}$ which
guarantees that the graviton has only two physical degrees of
freedom. This is precisely the definition of masslessness in $dS_4$ or 
$AdS_4$
space (for example see \cite{action} and references therein) and it should be stressed that the gravitons do not
have null cone propagation because the above action is not Weyl
invariant \cite{deser}. 

The propagator of the above spin-2 massless field can be written in the form:
\be
G^{0}_{\mu\nu;\mu'\nu'}(x,y)=(\partial_{\mu}\partial_{\mu'}u\partial_{\nu}\partial_{\nu'}u+\partial_{\mu}\partial_{\nu'}u\partial_{\nu}\partial_{\mu'}u)G^{0}(u)+g_{\mu\nu}g_{\mu'\nu'}E^{0}(u)+D[\cdots]
\label{mlesspr}
\ee
where $\partial_{\mu}={\partial \over \partial x^{\mu}}$,
$\partial_{\mu'}={\partial \over \partial y^{\mu'}}$. The  last term, denoted $D[\cdots]$, is a total derivative and drops out of the
calculation when integrated with a conserved energy momentum tensor. Thus, all physical information is encoded in the first two terms.

The process of finding the functions $G^{0}$ and $E^{0}$ is quite
complicated and is the result of solving a system of six coupled differential
equations \cite{mless}. We will  present here only the differential equation
that $G^{0}$ satisfies to show the difference between $AdS_4$ and
$dS_4$ space. This equation results from various integrations and has
the general form:
\be
u(u+2)G^{0}(u)''+4(u+1)G^{0}(u)'=C_1 +C_2 u
\label{diff}
\ee
where the constants $C_1$ and $C_2$ are to be fixed by the boundary
conditions. For the case of the $AdS_4$ space \cite{mless}, these
constants were set to zero so that the  $G^{0}$ function vanishes at
the boundary  at infinity ($u \rightarrow \infty$). Using the same condition also for the  $E^{0}$ function, the 
exact form of them was found to be:
\ba
G^{0}(u)&=&\frac{1}{8 \pi^2 H^2}\left[\frac{2(u+1)}{u(u+2)}-\log \frac{u+2}{u}\right]\cr
E^{0}(u)&=&-\frac{ H^2}{8
\pi^2}\left[\frac{2(u+1)}{u(u+2)}+4(u+1)-2(u+1)^2\log \frac{u+2}{u}\right]
\ea

For the case of the $dS_4$ space we iterated the procedure of
Ref.\cite{mless} imposing the condition \cite{allen} that the  $G^{0}$ and $E^{0}$
functions should be non-singular at the antipodal point
($u=-2$). The constants $C_1$ and $C_2$ were kept non-zero and played
a crucial role in finding a consistent solution. It is straightforward 
to find the full expression of these functions, but we only need to know their short distance behaviour. Then with this accuracy the answer is:
\ba
G^{0}(u)&=&-\frac{1}{8 \pi^2 H^2}\left[\frac{1}{u}+\log (-u)\right] +
\cdots \cr
E^{0}(u)&=&\frac{H^2}{8
\pi^2}\left[\frac{1}{u}+2(u+1)^2\log (-u)\right]+\cdots 
\ea

If we define $\Pi^{0}(u)=\frac{1}{H^4}\frac{E^{0}(u)}{G^{0}(u)}$, then 
for short distances ($H^2x^2 \ll 1$) where $u \rightarrow 0$ we get:
\ba
g_{\mu\nu}g_{\mu'\nu'}&\rightarrow & \delta_{\mu\nu} \delta_{\mu'\nu'} 
\cr
\partial_{\mu}\partial_{\nu'}u &\rightarrow& \mp H^2 \delta_{\mu \nu'} \cr G^0(u) &\rightarrow& {1 \over 4 \pi^2 H^4 \mu^2} \cr
\Pi^{0}(u) &\rightarrow& -1
\ea
and so we recover
the short distance limit of the massless flat Euclidean space propagator:
\be
G^{0}_{\mu\nu;\mu'\nu'}(x,y)=\frac{1}{4 \pi^2
\mu^2}(\delta_{\mu\mu'}\delta_{\nu\nu'}+\delta_{\mu\nu'}\delta_{\nu\mu'}-\delta_{\mu\nu}\delta_{\mu'\nu'})+\cdots
\label{mlessprop}
\ee
Of course this is just as expected.

In order to describe a spin-2 massive field it is necessary to add to the above 
action  a Pauli-Fierz mass term:
\be
\frac{S_m}{2 M^2}=\frac{S_0}{2 M^2}- \frac{m^2}{4}\int d^4 x \sqrt{g^0}(h_{\mu \nu}h^{\mu \nu}-h^2)
\ee
By adding this term we immediately lose the gauge invariance
associated with the $dS_4$ or $AdS_4$ symmetry group and the
massive gravitons acquire five degrees of freedom.

The propagator of this massive spin-2 field  can again be 
written in the form:
\be
G^{m}_{\mu\nu;\mu'\nu'}(x,y)=(\partial_{\mu}\partial_{\mu'}u\partial_{\nu}\partial_{\nu'}u+\partial_{\mu}\partial_{\nu'}u\partial_{\nu}\partial_{\mu'}u)G^{m}(u)+g_{\mu\nu}g_{\mu'\nu'}E^{m}(u)+D[\cdots]
\label{mivepr}
\ee
The last term
of the propagator in eq. (\ref{mivepr}), denoted $D[\cdots]$, is again a total derivative and thus
drops out of the calculation when integrated with a conserved $T_{\mu
\nu}$.

At his point we should emphasize that in case of an arbitrary massive
spin-2 field, the absence of
gauge invariance means that there is no guarantee that the field will
couple to a conserved current. However, in the context of a higher
dimensional theory  whose symmetry group is spontaneously broken by
some choice of vacuum metric,  the massive spin-2 graviton KK states 
couple to a conserved $T_{\mu \nu}$. One can understand this by the
following example. Consider the case of the most simple KK
theory, the one with one compact extra dimension. By the time 
we choose a vacuum metric \textit{e.g.} $g^0_{MN}={\rm diag}\left(\eta_{\mu \nu},1\right)$, the higher dimensional
symmetry is broken. If we denote the
graviton fluctuations around
the background metric by $h_{\mu \nu}$, $h_{\mu 5}$ and $h_{55}$, there is still the gauge freedom:
\ba
\delta h_{\mu \nu}&=&\partial_{\mu}\xi_{\nu}+\partial_{\nu}\xi_{\mu}\nonumber\\
\delta h_{\mu 5}&=&\partial_{\mu}\xi_{5}+\partial_{5}\xi_{\mu}\label{gauge}\\
\delta h_{5 5}&=&2\partial_{5}\xi_{5}\nonumber
\ea

If we Fourier decompose these fields, their $n$-th Fourier mode acquires
a mass $m_n \propto n$ with $n=0,1,2,\dots$, but there is mixing
between them. This means for example that $h^{(n)}_{\mu \nu}$ is not a 
massive spin-2 eigenstate \textit{etc.}. However, we can exploit the gauge
transformations (\ref{gauge}) to gauge away the massive $h^{(n)}_{\mu
5}$ and  $h^{(n)}_{55}$ and construct a pure spin-2 field (see for
example \cite{cho} and references therein). For a comprehensive account of KK theories see \cite{KK}. The new massive spin-2 field $\rho^{(n)}_{\mu \nu}$ is
invariant under (\ref{gauge}) and so its Lagrangian does not exhibit a
gauge invariance of the form $\delta \rho_{\mu
\nu}=\partial_{\mu}\chi_{\nu}+\partial_{\nu}\chi_{\mu}$. However,
since is originates from a Lagrangian that has the gauge invariance
(\ref{gauge}), it is bound to couple to a conserved $T_{\mu
\nu}$. The argument goes on for more complicated choices of vacuum
metric as for example warped metrics which are recently very popular
in brane-world constructions.

Again the functions $G^{m}$ and $E^{m}$ result from a complicated system of differential
equations \cite{mive}. In that case, the differential equation
that $G^{m}$ satisfies is:
\be
u(u+2)G^{m}(u)''+4(u+1)G^{m}(u)'\mp \left({m\over H}\right)^2 G^{m}(u)=C_1 +C_2 u
\label{diffm}
\ee
where the constants $C_1$ and $C_2$ are to be fixed by the boundary
conditions. For the case of the $AdS_4$ space \cite{mive}, these
constants were set to zero so that the  $G^{0}$ function vanishes at
the boundary  at infinity. Imposing 
additionally  the condition of fastest falloff at infinity ($u
\rightarrow \infty$) \cite{allen}, the exact form of the  $G^{m}$ and  $E^{0}$
function was found to be:
\ba
G^{m}(u)&=&\frac{\Gamma(\Delta)\Gamma(\Delta-1)}{16 \pi^2
\Gamma(2\Delta-2)
H^2}\left(\frac{2}{u}\right)^{\Delta}F(\Delta,\Delta-1,2\Delta-2,-{2\over 
u})\cr
E^{m}(u)&=&-\frac{2}{3}~\frac{\Gamma(\Delta-1) H^2}{16
\pi^2 \Gamma(2\Delta
-2)[2+(m/H)^2]}\left(\frac{2}{u}\right)^{\Delta}\times\cr
& \times & \left\{ \renewcommand{\arraystretch}{1.5} \begin{array}{l} \phantom{-} 3[2+(m/H)^2]
\Gamma(\Delta-2) u^2 F(\Delta-1,\Delta-2,2\Delta-2,-{2\over u})
\\-3(u+1)u F(\Delta-1,\Delta-1,2\Delta-2,-{2\over u})\\+[3+(m/H)^2]\Gamma(\Delta)F(\Delta,\Delta-1,2\Delta-2,-{2\over u}) \end{array} \right\}
\ea
where $\Delta={3\over 2}+{1\over 2}\sqrt{9+4(m/H)^2}$. 

For the case of the $dS_4$ space we iterated the procedure of
Ref.\cite{mive} imposing the condition \cite{allen} that the  $G^{m}$ and $E^{m}$
functions should be non-singular at the antipodal point
($u=-2$) and also finite as $m \rightarrow 0$. Again we kept the
constants $C_1$ and $C_2$  non-zero to obtain a consistent solution. It is straightforward to find the full expression of these functions, but we only need to know their short distance behaviour. Then with this accuracy the answer is:
\ba
G^{m}(u)&=&\frac{\Gamma(\Delta)\Gamma(3-\Delta)}{16 \pi^2
H^2}\left[F(\Delta,3-\Delta,2,{u+2\over 2})-1\right]+\cdots \cr
E^{m}(u)&=&-\frac{2}{3}~\frac{\Gamma(\Delta)\Gamma(3-\Delta) H^2}{4
\pi^2 [2-(m/H)^2]}~\times\cr
& \times & \left\{ \renewcommand{\arraystretch}{1.5} \begin{array}{l} -3[2-(m/H)^2]\left[{2(u+2)\over(\Delta-1)(\Delta-2)}F(\Delta-1,2-\Delta,2,{u+2\over 2})+{u(u+2)\over2}\right]
\\-3(u+1)\left[{2\over(\Delta-1)(\Delta-2)} F(\Delta-1,2-\Delta,1,{u+2\over 2})+(u+1)\right]
\\+[3-(m/H)^2]\left[F(\Delta,3-\Delta,2,{u+2\over 2})-1\right]
\end{array} \right\}+\cdots \phantom{aaa}
\ea
where $\Delta={3\over 2}+{1\over 2}\sqrt{9-4(m/H)^2}$. 

If we define
$\Pi^{m}(u)=\frac{1}{H^4}\frac{E^{m}(u)}{G^{m}(u)}$, then 
for short distances ($H^2x^2 \ll 1$) where $u \rightarrow 0$ we get:
\ba
G^m(u) &\rightarrow& {1 \over 4 \pi^2 H^4 \mu^2} \cr
\Pi^m(u) &\rightarrow& -{2\over 3}~\frac{~~3 \pm \left( {m \over H}
\right)^2}{~~2 \pm \left( {m \over H} \right)^2}
\label{ratio}
\ea

It is interesting to consider two massless flat limits. In the first
one  $m \rightarrow 0$ and $H \rightarrow 0$  while ~$m/H \rightarrow
\infty$. In this case, from
eq. (\ref{ratio}) we see that we
recover the Euclidean propagator for a massive graviton in flat space:
\be
G^{m}_{\mu\nu;\mu'\nu'}(x,y)=\frac{1}{4 \pi^2
\mu^2}(\delta_{\mu\mu'}\delta_{\nu\nu'}+\delta_{\mu\nu'}\delta_{\nu\mu'}-{2\over 
3}\delta_{\mu\nu}\delta_{\mu'\nu'})+\cdots
\ee
This is in agreement with the van Dam - Veltman - Zakharov
theorem. The second limit has $m \rightarrow 0$ and $H \rightarrow 0$
but $m/H \rightarrow 0$. In this case the propagator passes smoothly
to the one of the flat massless
case (\ref{mlessprop}):
\be
G^{0}_{\mu\nu;\mu'\nu'}(x,y)=\frac{1}{4 \pi^2
\mu^2}(\delta_{\mu\mu'}\delta_{\nu\nu'}+\delta_{\mu\nu'}\delta_{\nu\mu'}-\delta_{\mu\nu}\delta_{\mu'\nu'})+\cdots
\ee
This is in contrary to the van Dam - Veltman  - 
Zakharov discontinuity in flat
space.

In general, we may consider the limit with $m/H$ finite. Then for
small   $m/H$ the contribution to the $\delta_{\mu\nu}\delta_{\mu'\nu'}$ 
structure is  $ - 1  \pm m^2/6 H^2$.  Since observations agree to $1\%$ 
accuracy with the prediction of Einstein gravitational theory for the
bending of light by the sun, we obtain the limit ${m
\over H} \la 0.1$.

\section{Conclusions}

In summary, in this paper we showed that, by considering physics in
$dS_{4}$ or$AdS_{4}$ spacetime, 
one can circumvent the van Dam - Veltman - Zakharov 
theorem about non-decoupling of  the extra polarization states of a massive
graviton. It is shown that the smoothness of the $m\rightarrow{0}$
limit is ensured if the $H$ (``Hubble'') parameter, associated with
the horizon of $dS_{4}$ or
$AdS_{4}$ space, tends to zero slower than $m$. The above  requirement
can be  realized in various models and an interesting example
will be given elsewhere \cite{++}. Furthermore, if we keep $m/H$
finite, we can obtain models where massive
gravitons contribute to 
gravity and still have acceptable and interesting
phenomenology. Gravity will then be modified at all scales with
testable differences from the Einstein theory in future higher
precision observations. However, the dramatic modifications of gravity at large
distances that ``multigravity'' models suggest, will be hidden by the existence of the horizon which will
always be well before the scales that modifications would become
relevant. It will be interesting to investigate if the above  smooth limit can also
be obtained in the case of non-maximally symmetric spacetimes
(\textit{e.g.} considering FRW or Schwarzschild backgrounds) when the
inverse of  the graviton mass is much bigger than the characteristic curvature radius. This
issue is important for the phenomenology of ``multigravity'' and will
be addressed in other publication \cite{FRW}.

\textbf{Acknowledgments:} We would like to thank 
 Tibault Damour, Ruth Gregory, Panagiota Kanti, Keith A. Olive and
Maxim Pospelov for useful discussions. We are
indebted to Graham G. Ross for very important and stimulating
discussions and for careful reading of the manuscript. S.M.'s work is supported by the Hellenic State
Scholarship Foundation (IKY) \mbox{No. 
8117781027}. A.P.'s work is supported by the Hellenic State Scholarship
Foundation (IKY) \mbox{No. 8017711802}. This work   is
supported in part by the PPARC rolling grant PPA/G/O/1998/00567, by
the EC TMR grants  HRRN-CT-2000-00148 and  HPRN-CT-2000-00152.

\textbf{Addendum:} One day after this work had appeared in the hep-archives, ref.
\cite{POR} appeared, reaching the same result as this paper (for the
$AdS_{4}$) with a different approach. From this paper we became aware of
ref. \cite{HIG} where the smoothness of the limit of the graviton
propagator in $dS_{4}$ had been shown.


\begin{thebibliography}{99}



\bibitem{bigravity}  I.I. Kogan, S. Mouslopoulos, A. Papazoglou, G.G. Ross and
J. Santiago, Nucl. Phys. \textbf{B584} (2000) 313;\newline
S. Mouslopoulos and A. Papazoglou, JHEP \textbf{0011} (2000) 018.


\bibitem{GRS}  R. Gregory, V. A. Rubakov and S. M. Sibiryakov,
Phys. Rev. Lett. \textbf{84} (2000) 5928.


\bibitem{KR}  I.I. Kogan and G.G. Ross, Phys. Lett. \textbf{B485} (2000) 255.


\bibitem{multi}  I.I. Kogan, S. Mouslopoulos, A. Papazoglou and G.G. Ross, hep-th/0006030, to appear in Nucl. Phys. \textbf{B}.


\bibitem{reso}  C. Cs\'aki, J. Erlich and T.J. Hollowood,
Phys. Rev. Lett. \textbf{84} (2000) 5932.


\bibitem{dvali1} G. Dvali, G. Gabadadze and M. Porrati,
Phys. Lett. \textbf{B484} (2000) 112.



\bibitem{cmb} P. Bin\'etruy and J. Silk, astro-ph/0007452.




\bibitem{VZ}  H. van Dam and M. Veltman, Nucl. Phys. \textbf{B22} (1970) 397; 
\newline
V.I. Zakharov, JETP Lett. \textbf{12} (1970) 312.


\bibitem{ghost} G. Dvali, G. Gabadadze and M. Porrati,
Phys. Lett. \textbf{B484} (2000) 129;\newline L. Pilo, R. Rattazzi and
A. Zaffaroni, JHEP \textbf{0007} (2000) 056.


\bibitem{weak}  D.Z.Freedman, S.S.Gubser, K.Pilch and N.P.Warner,
hep-th/9904017;\newline E.Witten, hep-ph/0002297.



\bibitem{rest}
C. Csaki, J. Erlich and T. J. Hollowood, Phys. Lett. \textbf{B481}
(2000) 107;\newline
R. Gregory, V. A. Rubakov and S. M. Sibiryakov, Phys. Lett. \textbf{B489}
(2000) 203.


\bibitem{olive} P. Kanti, I.I. Kogan, K.A. Olive and M. Pospelov,
Phys. Rev. \textbf{D61} (2000) 106004.




\bibitem{++}  I.I. Kogan, S. Mouslopoulos and A. Papazoglou, hep-th/0011141.



\bibitem{HIG}  A.Higuchi, Nucl. Phys. \textbf{B282} (1987) 397;
\newline
A.Higuchi, Nucl. Phys. \textbf{B325} (1989) 745.


\bibitem{action} I.L. Buchbinder, D.M. Gitman and V.D. Pershin,
Phys. Lett. \textbf{B492} (2000) 161.


\bibitem{deser} S. Deser and R.I. Nepomechie, Ann. Phys. \textbf{154}
(1984) 396.


\bibitem{mless} E. D'Hoker, D.Z. Freedman, S.D. Mathur, A. Matusis and 
L. Rastelli, Nucl. Phys. \textbf{B562} (1999) 330.


\bibitem{allen} B. Allen and T. Jacobson,
Commun. Math. Phys. \textbf{103} (1986) 669.


\bibitem{cho} Y.M. Cho and S.W. Zoh, Phys. Rev. \textbf{D46} (1992) 2290.



\bibitem{KK}  \textit{``Modern Kaluza-Klein Theories''}, edited by T.
Appelquist, A. Chodos and P.G.O. Freund, Addison-Wesley, (1987).


\bibitem{mive} A. Naqvi, JHEP \textbf{9912} (1999) 025.



\bibitem{FRW}  I.I. Kogan, S. Mouslopoulos, A. Papazoglou and
G.G. Ross, work in progress.


\bibitem{POR}  M. Porrati, hep-th/0011152.


\end{thebibliography}
\end{document}